\documentstyle{article}

\begin{document}

\title{First-principles calculations of the electronic structure and spectra of
strongly correlated systems:dynamical mean-field theory}
\author{V.I. Anisimov, A.I.Poteryaev, M.A.Korotin, A.O.Anokhin \\
Institute of Metal Physics, Ekaterinburg,GSP-170,Russia \\
\\
G.Kotliar \\
Serin Physics Laboratory,\\
Rutgers University, Piscataway, New Jersey 08854, USA}
\date{}
\maketitle

\begin{abstract}
A recently developed dynamical mean-field theory in the iterated
perturbation theory approximation was used as a basis for construction of
the "first principles" calculation scheme for investigating electronic
structure of strongly correlated electron systems. This scheme is based on
Local Density Approximation (LDA) in the framework of the Linearized
Muffin-Tin-Orbitals (LMTO) method. The classical example of the doped
Mott-insulator La$_{1-x}$Sr$_x$TiO$_3$ was studied by the new method and the
results showed qualitative improvement in agreement with experimental
photoemission spectra.
\end{abstract}

\section{Introduction}

The accurate calculation of the electronic
structure of materials starting from first principles is 
a
challenging  problem
in condensed matter science
since unfortunately, except for small molecules, it is impossible to
solve many-electron problem without severe approximations. 

For materials where the  kinetic energy of the electrons
is more important than the 
Coulomb interactions,
the most successful first principles method is the Density Functional theory (DFT)
within the Local (Spin-) Density Approximation (L(S)DA)\cite{lda}, where the
many-body problem is mapped into a non-interacting system with a
one-electron exchange-correlation potential approximated by that of the
homogeneous electron gas.

It is by now,  generally accepted that the spin density functional theory in
the local approximation   is a reliable
starting point   for first principle calculations
of material properties of  weakly correlated solids (For a review see \cite{RMP}).
The situation is very different 
when we consider
more strongly correlated materials, (systems containing
f and d electrons).
In a very simplified  view LDA can be regarded as a Hartree-Fock
approximation with orbital-independent (averaged) one-electron potential.
This approximation is very crude for strongly correlated systems,
where the on-cite Coulomb interaction between d- (or f-) electrons of
transition metal (or rare-earth metal) ions (Coulomb parameter $U$) is
strong enough to overcome kinetic energy which is of the order of band width 
$W$. In the result LDA gives qualitatively wrong answer even for such simple
systems as Mott insulators with integer number of electrons per cite
(so-called "undoped Mott insulators") . For example insulators CoO and La$%
_2$CuO$_4$ are predicted to be metallic by LDA.

The search for a "first principle"  computational  scheme
of physical properties of strongly correlated electron systems
which is as successful as the LDA in weakly correlated systems,
is highly desirable given the considerable importance of this
class of materials
and is a subject of intensive current research.
Notable examples  of first principle schemes that have been
applied to srongly correlated electron systems are the LDA+U method
\cite{ldau} which is akin to 
orbital-spin-unrestricted Hartree-Fock method using a basis
of  LDA   wave functions, ab initio unrestricted Hartree Fock
calculations \cite{HF}  and the use of 
constrained LDA  to derive model parameters of model hamiltonians which are then treated by
exact diagonalization of small clusters or other approximations \cite{hybertsen}.

 Many interesting effects, such as orbital and charge
ordering in  transition metal compounds were successfully described by LDA+U
method \cite{review}.
However for strongly correlated metals Hartree-Fock
approximation is too crude and more sophisticated approaches are needed.

Recently the dynamical mean-field theory was developed \cite{kotliar} which
is based on the mapping of lattice models onto quantum impurity models
subject to a self-consistency condition. The resulting impurity model can be
solved by various approaches (Quantum Monte Carlo, exact diagonalization)
but the most promising for the possible use in "realistic" calculation
scheme is Iterated Perturbation Theory (IPT) approximation, which was proved
to give results in a good agreement with more rigorous methods.

This paper is the first in a series where we plan to integrate
recent develompements of  the dynamical mean 
field approach with state of the art
band structure calculation techniques to generate an "ab initio"
scheme for the calculation of the electronic structure of
correlated solids.
For a review of the historical development of the dynamical
mean field approach in its various implementations see 
ref \cite{kotliar}
.
In this paper we implement  the dynamical mean-field theory
in the iterated perturbation theory approximation,
and carry out the band structure calculations using a LMTO
basis .
The calculational scheme is described in section 2
.  We present results obtained applying this method
to  La$_{1-x}$Sr$_x$TiO$_3$ which
is a classical example of strongly correlated metal.

\section{The calculation scheme}

~\qquad In order to be able to implement the achievements of Hubbard model
theory to LDA one needs the method which could be mapped on tight-binding
model. The Linearized Muffin-Tin Orbitals (LMTO) method in orthogonal
approximation \cite{lmto} can be naturally presented as tight-binding
calculation scheme (in real space representation):

\begin{equation}
H_{LMTO}=\sum\limits_{ilm,jl^{\prime }m^{\prime },\sigma }(\delta
_{ilm,jl^{\prime }m^{\prime }}\,\epsilon _{il}\,\widehat{n}_{ilm\sigma
}+t_{ilm,jl^{\prime }m^{\prime }}\widehat{\,c}_{ilm\sigma \,}^{\dagger }%
\widehat{c}_{jl^{\prime }m^{\prime }\sigma })
\end{equation}

($i$ - site index, $lm$ - orbital indexes).

As we have mentioned above, LDA one-electron potential is orbital -
independent and Coulomb interaction between d-electrons is taken into
account in this potential in an averaged way. In order to generalize this
Hamiltonian by including Coulomb correlations, one must add interaction term:

\begin{equation}
H_{int}=\frac 12\sum_{ ilmm^{\prime }\sigma \sigma ^{\prime }  \\ %
m\sigma \neq m^{\prime }\sigma ^{\prime }  } U_{il}\widehat{\,n}%
_{ilm\sigma }\,\widehat{n}_{ilm^{\prime }\sigma ^{\prime }}
\end{equation}

We neglected for a while exchange terms and dependence of Coulomb parameter $%
U$ on the particular pair of orbitals $mm^{\prime }$. Through the following
we will assume that only for one shell $l_d$ of one type of atoms $i_d$ (for
example d-orbitals of the transition metal ions) Coulomb interaction will be
taken into account ($U_{il}=U\delta _{il,i_dl_d}$), and in the following
iesndex $il$ will be omitted. All other orbitals will be considered as
resulting in the itinerant bands and well described by LDA. Such separation
of the electronic states into localized and itinerant is close in spirit to
the Anderson model.

To avoid double-counting one must in the same time subtract the averaged
Coulomb interaction energy term, which we assume is present in LDA.
Unfortunately there is no direct connection between Hubbard model and LDA
(because LDA is based on the homogeneous electron gas theory and not on the
localized atomic type orbitals representation) and it is impossible to
express rigorously LDA-energy through d-d Coulomb interaction parameter $U$.
However it is known that LDA total energy as a function of the total number
of electrons is a good approximation and the value of the Coulomb parameter $%
U$ obtained in LDA calculation agrees well with experimental data and
results of the more rigorous calculations \cite{Ucalc}.That leads us to the
suggestion that a good approximation for the LDA part of the Coulomb
interaction energy will be:

\begin{equation}
E_{Coul}=\frac 12Un_d(n_d-1)
\end{equation}
($n_d=\sum\limits_{m\sigma }n_{m\sigma }$ total number of d-electrons).

In LDA-Hamiltonian $\epsilon _{d}$ has a meaning of the
LDA-one-electron eigenvalue for d-orbitals. It is known that in LDA
eigenvalue is the derivative of the total energy over the occupancy of the
orbital: 
\begin{equation}
\epsilon _{d}=\frac d{dn_d}E_{LDA}
\end{equation}

If we want to introduce new $\epsilon _d^0$ where d-d Coulomb interaction is
excluded we must define them as: 
\begin{equation}
\epsilon _d^0=\frac d{dn_d}(E_{LDA}-E_{Coul})=\epsilon _{d%
}-U(n_d-\frac 12)
\end{equation}

Then new Hamiltonian will have the form: 
\begin{eqnarray}
H &=&H^0+H_{int}  \nonumber \\
H^0 &=&\sum\limits_{ilm,jl^{\prime }m^{\prime },\sigma }(\delta
_{ilm,jl^{\prime }m^{\prime }}\,\epsilon _{il}^0\,\widehat{n}_{ilm\sigma
}+t_{ilm,jl^{\prime }m^{\prime }}\widehat{\,c}_{ilm\sigma \,}^{\dagger }%
\widehat{c}_{jl^{\prime }m^{\prime }\sigma })
\end{eqnarray}

In reciprocal space matrix elements of the operator $H^0$ are: 
\begin{equation}
H_{qlm,q^{\prime }l^{\prime }m^{\prime }}^0({\bf k})=H_{qlm,q^{\prime
}l^{\prime }m^{\prime }}^{LDA}({\bf k})-\delta _{qlm,q^{\prime }l^{\prime
}m^{\prime }}\delta _{ql,i_dl_d}U(n_d-\frac 12)
\end{equation}

($q$ is an index of the atom in the elementary unit cell).

In the dynamical mean-field theory the effect of Coulomb correlation is
described by self-energy operator in local approximation. The Green function
is: 
\begin{equation}
G_{qlm,q^{\prime }l^{\prime }m^{\prime }}(i\omega )=\frac 1{V_B}\int d{\bf %
k\,[}i\omega +\mu -H_{qlm,q^{\prime }l^{\prime }m^{\prime }}^0({\bf k}%
)-\delta _{qlm,q^{\prime }l^{\prime }m^{\prime }}\delta _{ql,i_dl_d}\Sigma
(i\omega )]^{-1}
\end{equation}
($[...]^{-1}$ means inversion of the matrix, integration is over Brillouin
zone, $\mu $ is chemical potential, $V_B$ is a volume of Brillouin zone).

In the following we will consider paramagnetic case, orbital and spin
degenerate system, so that self-energy $\Sigma (i\omega )$ does not depend
on orbital and spin indexes. One can define effective Anderson model Green
function through: 
\begin{equation}
G(i\omega )=G_{i_dl_dm,i_dl_dm}(i\omega )=(i\omega +\mu -\Delta (i\omega
)-\Sigma (i\omega ))^{-1}
\end{equation}
where $\Delta (i\omega )$ is effective impurity hybridization function. The
effective medium "bath" Green function $G^0$ is defined as: 
\begin{equation}
G^0(i\omega )=(i\omega +\widetilde{\mu }-\Delta (i\omega
))^{-1}=(G^{-1}(i\omega )+\Sigma (i\omega )+\widetilde{\mu }-\mu )^{-1}
\end{equation}
($\widetilde{\mu }$ is chemical potential of the effective medium).

The chemical potential of the effective medium $\widetilde{\mu }$ is varied
to satisfy Luttinger theorem condition: 
\begin{equation}
\frac 1\beta \sum\limits_{i\omega _n}e^{i\omega _n0^{+}}G(i\omega _n)\frac
d{d(i\omega _n)}\Sigma (i\omega _n)=0
\end{equation}

In iterated perturbation theory approximation the {\it anzatz} for the
self-energy is based on the second order perturbation theory term calculated
with "bath" Green function $G^0$: 
\begin{equation}
\Sigma ^0(i\omega _s)=-(N-1)U^2\frac 1{\beta ^2}\sum\limits_{i\omega
_n}\sum\limits_{ip_m}G^0(i\omega _m+ip_n)G^0(i\omega _m)G^0(i\omega _s-ip_n)
\label{Sigma0}
\end{equation}

$N$ is a degeneracy of orbitals including spin, $\beta =\frac 1{kT}$,
Matsubara frequencies $\omega _s=\frac{(2s+1)\pi }\beta ;p_n=\frac{2n\pi }%
\beta $ $;s,n$ integer numbers.

The term $\Sigma ^0$ is renormalized to insure correct atomic limit: 
\begin{equation}
\Sigma (i\omega )=Un(N-1)+\frac{A\Sigma ^0(i\omega )}{1-B\Sigma ^0(i\omega )}
\end{equation}
($n$ is orbital occupation $n=\frac 1\beta \sum\limits_{i\omega
_n}e^{i\omega _n0^{+}}G(i\omega _n)$), 
\begin{eqnarray}
B &=&\frac{U[1-(N-1)n]-\mu +\widetilde{\mu }}{U^2(N-1)n_0(1-n_0)} \\
A &=&\frac{n[1-(N-1)n]+(N-2)D[n]}{n_0(1-n_0)} \\
n_0 &=&\frac 1\beta \sum\limits_{i\omega _n}e^{i\omega _n0^{+}}G^0(i\omega
_n)
\end{eqnarray}
correlation function $D[n]\equiv <\widehat{n}\,\widehat{n}>_{CPA}$ is
calculated using Coherent Potential Approximation (CPA) for the Green
function with parameter $\delta \mu $ chosen to preserve orbital occupation $%
n$ : 
\begin{eqnarray}
G_{CPA}(i\omega ) &=&\frac{[1-n(N-1)]}{i\omega +\mu -\Delta (i\omega
)+\delta \mu }+\frac{n(N-1)}{i\omega +\mu -\Delta (i\omega )-U+\delta \mu }
\\
n &=&\frac 1\beta \sum\limits_{i\omega _n}e^{i\omega _n0^{+}}G_{CPA}(i\omega
_n) \\
D[n] &=&n\,\sum\limits_{i\omega _n}e^{i\omega _n0^{+}}\frac 1{i\omega +\mu
-\Delta (i\omega _n)-U+\delta \mu }
\end{eqnarray}

The Matsubara frequency convolution in (\ref{Sigma0}) was performed with
time variables representation using Fast Fourier Transform algorithm for
transition from energy to time variables and back: 
\begin{eqnarray}
G^0(\tau ) &=&\frac 1\beta \sum\limits_{i\omega _n}e^{-i\omega _n\tau
}G^0(i\omega _n) \\
\Sigma (\tau ) &=&-(N-1)U^2G^0(\tau )G^0(\tau )G^0(-\tau ) \\
\Sigma ^0(i\omega _n) &=&\int\limits_0^\beta d\tau \,e^{i\omega _n\tau
}\,\Sigma (\tau )
\end{eqnarray}

The serious problem is to perform integration in {\bf k}-space over
Brillouin zone. For this we used generalized Lambin-Vigneron algorithm \cite
{lambin}. We define new matrix $H({\bf k},z)$ as: 
\begin{equation}
H({\bf k},z)=H^0({\bf k)+}\Sigma (z)
\end{equation}
$z$ - complex energy, term $\Sigma (z)$ is added only to diagonal element of 
$H$-matrix corresponding to d-orbitals. In this matrix notations Green
function is : 
\begin{equation}
G(z)=\frac 1{V_B}\int d{\bf k[}z-H({\bf k},z)]^{-1}
\end{equation}
After diagonalization, $H({\bf k},z)$ matrix can be expressed through
diagonal matrix of its eigenvalues $D({\bf k},z)$ and eigenvectors matrix $U(%
{\bf k},z)$ : 
\begin{equation}
H({\bf k},z)=U({\bf k},z)D({\bf k},z)U^{-1}({\bf k},z)
\end{equation}
and Green function: 
\begin{equation}
G(z)=\frac 1{V_B}\int d{\bf k}U({\bf k},z)[z-D({\bf k},z)]^{-1}U^{-1}({\bf k}%
,z)
\end{equation}
A particular matrix element of Green function is calculated as: 
\begin{equation}
G_{ij}(z)=\sum\limits_n\frac 1{V_B}\int d{\bf k}\frac{U_{in}({\bf k}%
,z)U_{nj}^{-1}({\bf k},z)}{z-D_n({\bf k},z)}  \label{Gf1}
\end{equation}

In analytical tetrahedron method the irreducible wedge of the Brillouin zone
is divided into a set of tetrahedra and the total integral is calculated as
a sum over the tetrahedra. To perform integration over a given tetrahedron
with four corners at vectors ${\bf k}_i$ ($i=1,2,3,4$) denominator of the
fraction in equation (\ref{Gf1}) is interpolated as a linear function in 
{\bf k}-space. In the result the integral over one tetrahedron is expressed
through the values of numerator and denominator at the corners of the
tetrahedron: 
\begin{equation}
\sum\limits_n\frac 1{V_B}\int\limits_vd{\bf k}\frac{U_{in}({\bf k}%
,z)U_{nj}^{-1}({\bf k},z)}{z-D_n({\bf k},z)}=\sum\limits_n\sum%
\limits_{i=1}^4r_i^nU_{in}({\bf k}_i,z)U_{nj}^{-1}({\bf k}_i,z)\frac v{V_B}
\label{Gftetra}
\end{equation}
$v$ is tetrahedron volume 
\begin{eqnarray}
r_i^n &=&\frac{(z-D_n({\bf k}_i,z))^2}{\prod\limits_{k(\neq i)}(D_n({\bf k}%
_k,z)-D_n({\bf k}_i,z))}+ \\
&&\sum\limits_{j(\neq i)}\frac{(z-D_n({\bf k}_j,z))^3}{\prod\limits_{k(\neq
j)}(D_n({\bf k}_k,z)-D_n({\bf k}_j,z))}\frac{\ln [(z-D_n({\bf k}%
_j,z))/(z-D_n({\bf k}_i,z)]}{(D_n({\bf k}_i,z)-D_n({\bf k}_j,z))}  \nonumber
\end{eqnarray}

The self-energy $\Sigma (i\omega _n)$ and Green function $G(i\omega _n)$ are
calculated at the imaginary Matsubara frequencies $i\omega _n=i\pi
(2n+1)/\beta $ . It is enough to calculate expectation values, such as
orbital occupancies $n$ , but in order to calculate spectral properties one
need to know Green function on the real axis. The real axis equivalent of
equations (\ref{Sigma0}) is much more complicated and hard to implement
numerically than Matsubara frequencies version. It is much more convenient
to perform analytical continuation from imaginary energy values to the real
ones. For such continuation we have used Pad\'e approximant algorithm \cite
{pade}. If one has a set of the complex energies $z_i$ $(i=1,...,M)$ and the
set of values of the analytical function $u_i$ , then the approximant is
defined as continued fraction: 
\begin{equation}
C_M(z)=\frac{a_1}{1+}\frac{a_2(z-z_{2)}}{1+}...\frac{a_M(z-z_{M-1})}1
\end{equation}
where the coefficients $a_i$ are to be determined so that: 
\begin{equation}
C_M(z_i)=u_{i,}\ i=1,...,M
\end{equation}
The coefficients $a_i$ are then given by the recursion: 
\begin{eqnarray}
a_i &=&g_i(z_i),\ g_1(z_i)=u_i,\;i=1,...,M \\
g_p(z) &=&\frac{g_{p-1}(z_{p-1})-g_{p-1}(z)}{(z-z_{p-1})g_{p-1}(z)},\;p\geq 2
\end{eqnarray}
The recursion formula for continued fraction finally yields: 
\begin{equation}
C_M(z)=A_M(z)/B_M(z)
\end{equation}
where 
\begin{eqnarray}
A_{n+1}(z) &=&A_n(z)+(z-z_n)a_{n+1}A_{n-1}(z)  \nonumber \\
B_{n+1}(z) &=&B_n(z)+(z-z_n)a_{n+1}B_{n-1}(z)
\end{eqnarray}
and 
\[
A_0=0,\;A_1=a_1,\;B_0=B_1=1
\]

We have found that the most convenient way is to use analytical continuation
not for the Green function $G$ but only for self-energy $\Sigma $, and then
to calculate $G$ directly on the real axis through the Brillouin zone
integration (\ref{Gftetra}).

\section{Results}

~\qquad We have applied the above described calculation scheme to the doped
Mott insulator La$_{1-x}$Sr$_x$TiO$_3$. LaTiO$_3$ is a Pauli-paramagnetic
metal at room temperature and below T$_N$=125 K antiferromagnetic insulator
with a very small gap value ( 0.2 eV) . Doping by a very small value of Sr
(few percent) leads to the transition to paramagnetic metal with a large
effective mass. As photoemission spectra of this system also show strong
deviation from the noninteracting electrons picture, La$_{1-x}$Sr$_x$TiO$_3$
is regarded as an example of strongly correlated metal.

The crystal structure of LaTiO$_3$ is slightly distorted cubic perovskite.
The Ti ions have octahedral coordination of oxygen ions and $t_{2g}$-$e_g$
crystal field splitting of d-shell is strong enough to survive in solid. On
Fig.1 the total and partial DOS of paramagnetic LaTiO$_3$ are presented as
obtained in LDA calculations (LMTO method). On 3 eV above O2p-band there is
Ti-3d-band splitted on $t_{2g}$ and $e_g$ subbands which are well separated
from each other. Ti$^{4+}$-ions have d$^1$ configuration and $t_{2g}$ band
is 1/6 filled.

As only $t_{2g}$ band is partially filled and $e_g$ band is completely
empty, it is reasonable to consider Coulomb correlations between $t_{2g}-$%
electrons only and degeneracy factor $N$ in Eq. (\ref{Sigma0}) is equal 6.
The value of Coulomb parameter $U$ was calculated by the supercell procedure 
\cite{Ucalc} regarding only $t_{2g}-$electrons as localized ones and
allowing $e_g-$electrons participate in the screening. This calculation
resulted in a value 3 eV. As the localization must lead to the energy gap
between electrons with the same spin, the effective Coulomb interaction will
be reduced by the value of exchange parameter $J$=1 eV. So we have used
effective Coulomb parameter $U_{eff}$=2 eV. The results of the calculation
for x=0.06 (doping by Sr was immitated by the decreasing on x the total
number of electrons)
are presented in the form of the $t_{2g}$-DOS on Fig.2. Its general
form is the same as for model calculations: strong quasiparticle peak on the
Fermi energy and incoherent subbands below and above it corresponding to the
lower and upper Hubbard bands.

The appearance of the incoherent lower Hubbard band in our DOS leads to
qualitatively better agreement with photoemission spectra. On Fig.3 the
experimental XPS for La$_{1-x}$Sr$_x$TiO$_3$ (x=0.06) \cite{spectr}
is presented with 
non-interacting (LDA) and
interacting (IPT) occupied DOS broadened to imitate experimental resolution.
The main correlation effect: simultaneous presence of coherent and
incoherent band in XPS is successfully reproduced in IPT calculation.
However, as one can see, IPT overestimates the strength of the coherent
subband.

\section{Conclusions}

In this publication we described how one can interface 
methods for realistic band  structure calculations with
the recently developed dynamical mean field technique
 to  obtain a fully "ab initio" method for calculating the electronic
spectra of solids.

With respect to earlier calculations, this work introduces
several methodological advances: the dynamical mean field
equations are incorporated into a realistic electronic
structure calculation scheme, with
parameters obtained from a first principle calculation and with 
the realistic orbital degeneracy of the compound.

To check our method we applied to doped titanates for
which a large body of model calculation studies using 
dynamical mean field theory exists. There results are very encouraging considering 
the experimental uncertainties of the analysis of the photoemission
spectra of these compounds.

We have used two relative accurate (but still approximate ) methods for the
solution of the band structure aspect and the correlation aspects of
this problem: the LMTO in the ASA approximation and the IPT approximation.
In principle, one can use other techniques for handling these two
aspects of the problem and
further application to more complicated materials are necessary
to determine the  degree   of quantitative accuracy of the  method.

\newpage

\section{Figure captions}

~\quad Fig. 1. Noninteracting ($U$=0) total and partial density of states
(DOS) for LaTiO$_3.$

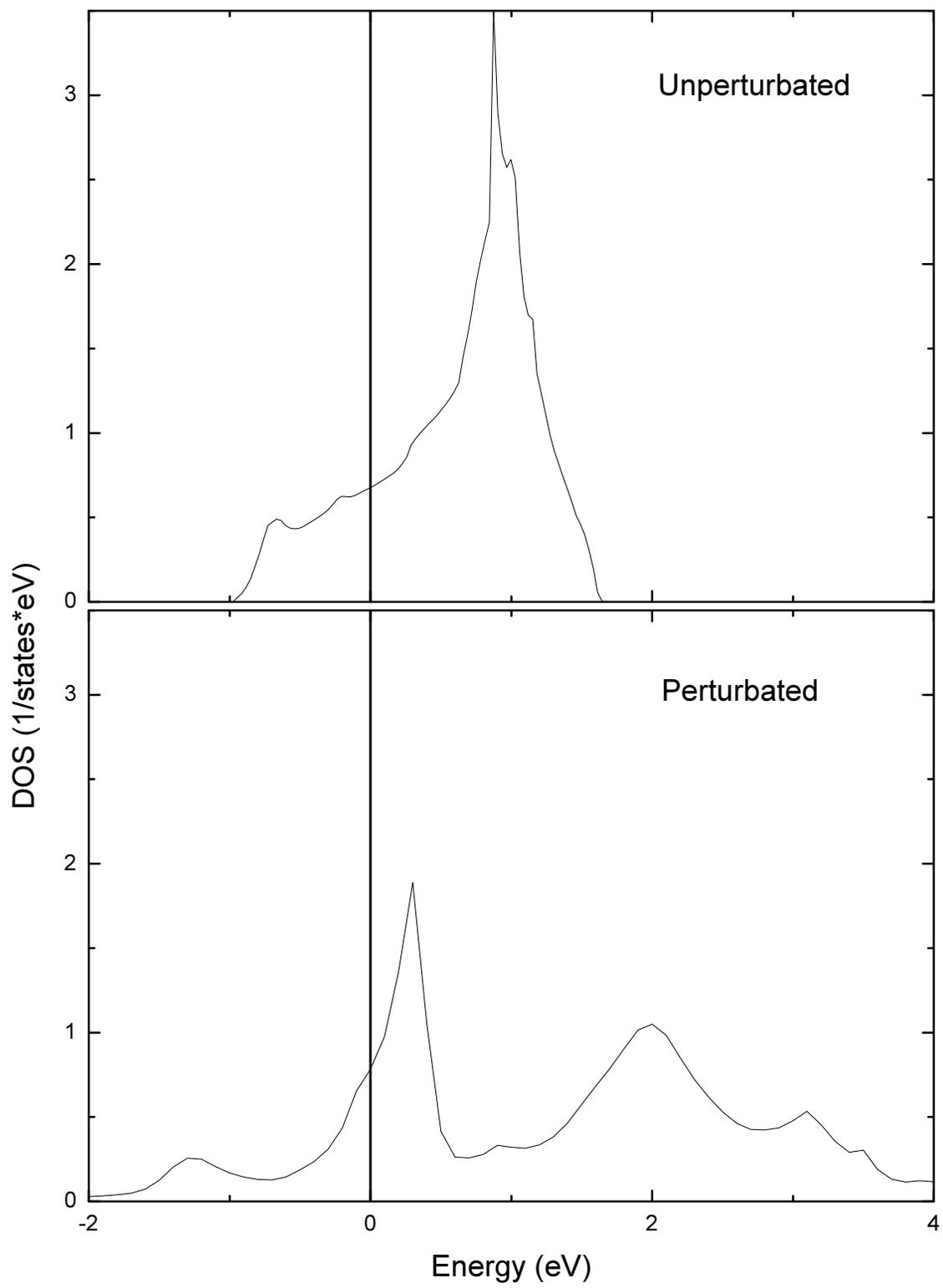
Fig. 2. Partial ($t_{2g})$ DOS obtained in IPT calculations in comparison
with noninteracting DOS.

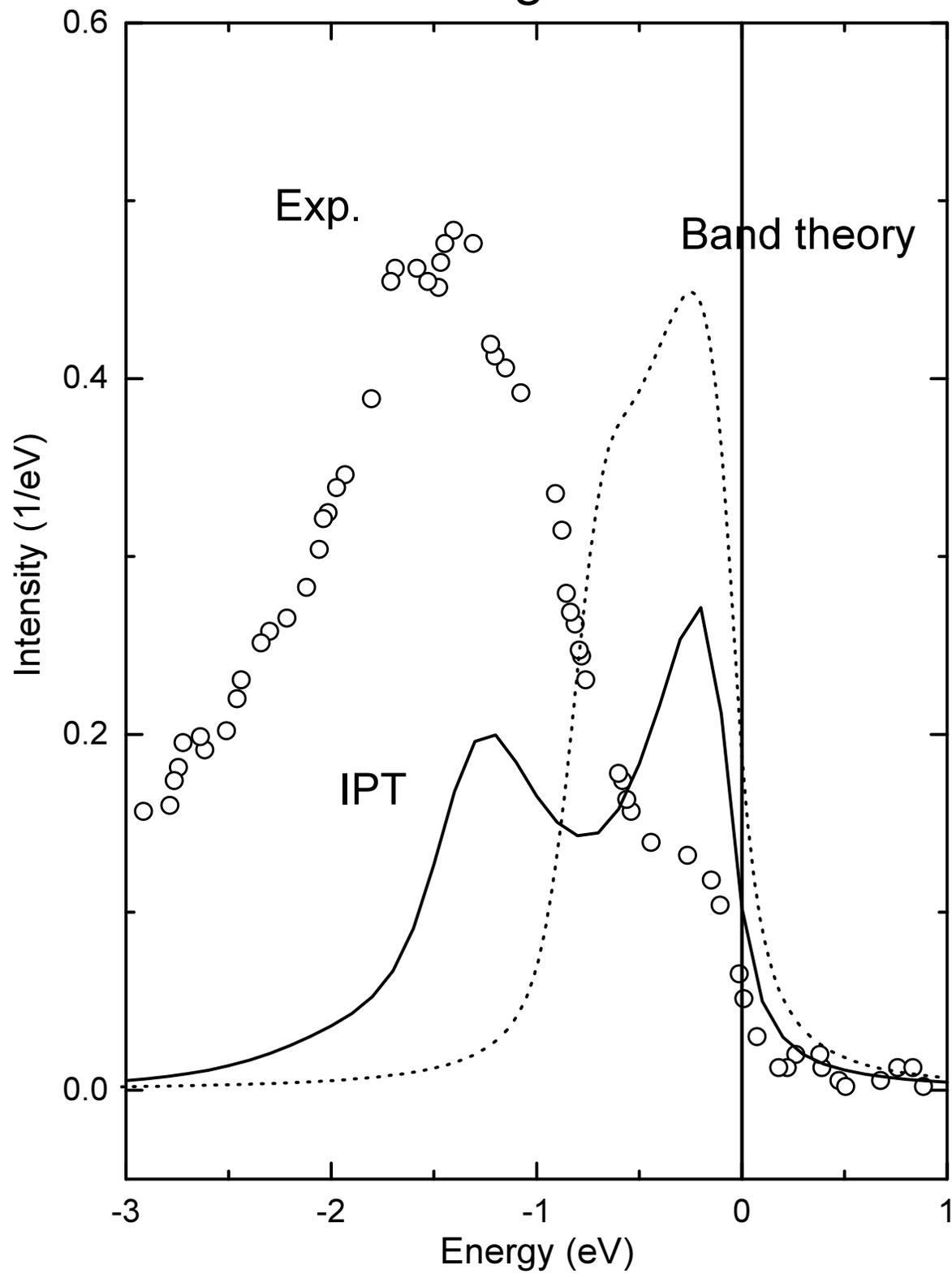
Fig. 3. Experimental and theoretical photoemission spectra of La$_{1-x}$Sr$_x$TiO$_3$ (x=0.06).

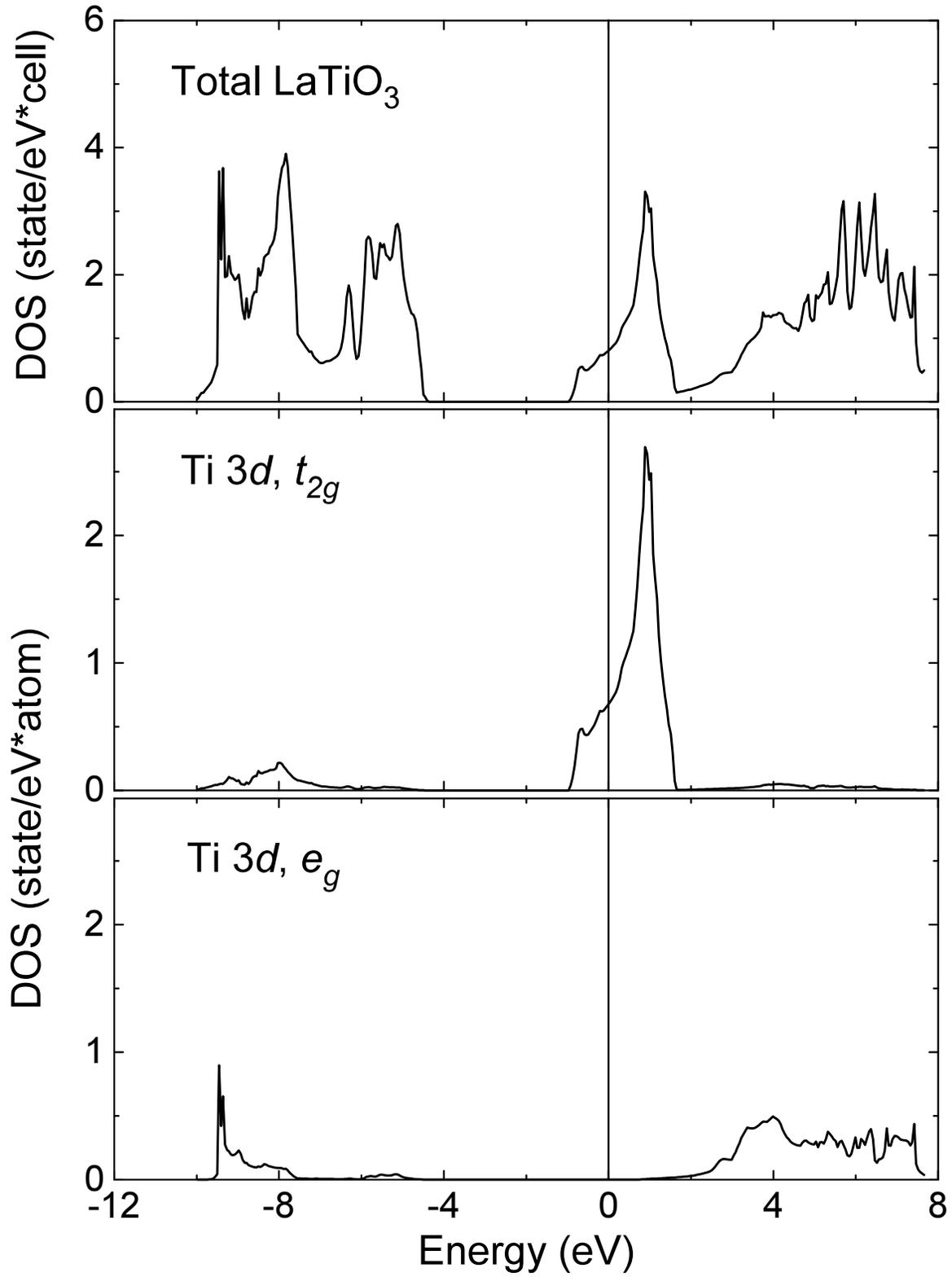
\end{document}